\documentclass[aps,prl,reprint,nobalancelastpage,nofootinbib, preprintnumbers]{revtex4-2}

\usepackage[utf8]{inputenc}

\usepackage{amsmath}
\usepackage{amssymb}
\usepackage[all]{xy}
\usepackage{tikz}
\usetikzlibrary{calc}

\usepackage[toc,page]{appendix}
\usepackage{enumerate}
\usepackage{tensor}
\usepackage{booktabs}
\usepackage{bbm}
\usepackage{youngtab}
\usepackage{slashed}
\usepackage{multirow}
\usepackage{makecell}
\usepackage{float}
\usepackage{comment}
\usepackage{verbatim}
\usepackage{xcolor}
\usepackage[framemethod=tikz]{mdframed}
\usepackage{soul}
\usepackage{mathtools}
\usepackage[inline]{enumitem}

\usepackage{amsthm}

\newcommand{\ab}{|}
\newcommand{\der}{\partial}
\newcommand{\de}{\mathrm{d}}
\newcommand{\dd}{\mathrm{d}}

\newcommand{\vevc}{\chi}
\newcommand{\sdil}{\phi}

\newcommand{\tdelta}{\tilde{\Phi}}

\newcommand{\tsigma}{\tilde{\sigma}}

\newcommand{\ttau}{\tilde{\tau}}

\newcommand{\e}{\mathrm{e}}

\newcommand{\p}{\mathrm{P}}

\newcommand{\tvec}[2]{\left(\!\!\begin{array}{c}
    #1 \\
    #2
\end{array}\!\!\right)}

\newcommand{\matr}[4]{\left(\!\begin{array}{cc}
    #1 \, & \, #2 \\
    #3 \, & \, #4
\end{array}\!\right)}

\usepackage{hyperref}
\hypersetup{
    colorlinks=true,
    linkcolor=purple,
    citecolor=purple,
    filecolor=purple,      
    urlcolor=purple,
}

\begin{document}
\numberwithin{equation}{section}

\preprint{UUITP-32/23}
\title{Connecting flux vacua through scalar field excursions}

\author{Gary Shiu}
\email{shiu@physics.wisc.edu}
\affiliation{Department of Physics, University of Wisconsin-Madison, 1150 University Avenue, Madison, WI 53706, USA}

\author{Flavio Tonioni}
\email{flavio.tonioni@kuleuven.be}
\affiliation{Instituut voor Theoretische Fysica, K.U. Leuven, Celestijnenlaan 200D, B-3001 Leuven, Belgium}

\author{Vincent Van Hemelryck}
\email{vincent.vanhemelryck@physics.uu.se}
\affiliation{Institutionen för Fysik och Astronomi, Uppsala Universitet, Box 803, SE-751 08 Uppsala, Sweden}

\author{Thomas Van Riet}
\email{thomas.vanriet@kuleuven.be}
\affiliation{Instituut voor Theoretische Fysica, K.U. Leuven, Celestijnenlaan 200D, B-3001 Leuven, Belgium}

\begin{abstract}
We show how flux vacua that differ from each other in flux quanta can be seen as different vacua in a single scalar potential of an enlarged field space, which resolves the separation by thin domain walls. This observation, which is motivated by the
AdS Distance Conjecture,
allows one to compute distances between different vacua using the usual field space metric. We verify for explicit examples such as scale-seperated IIA flux vacua and the IIB Freund-Rubin vacua that the Distance Conjecture (for scalar fields) is satisfied and that the asymptotic directions in the enlarged field space are indeed hyperbolic. 
This enlarged field space contains the tachyon fields on the unstable $\tilde{\mathrm{D}}p$-branes of type II string theory, which can induce the brane charges of the stable D-branes. We suggest that requiring continuous interpolations refines the Cobordism Conjecture and postdicts the existence of unstable $\tilde{\mathrm{D}}p$-branes.
\end{abstract}

\maketitle

\section{Introduction}
The rich structure of the vacuum manifold of string theory, a.k.a.~the Landscape, is often depicted by
an energy functional (scalar potential) with many local minima. While this picture is useful for conveying to the public, it is not commonly used in quantitative investigations of the Landscape. In particular, we do not tend to think of flux vacua which differ by flux quantum numbers or compactification manifold as different vacua in a single potential.
Instead, vacua that differ in flux numbers or choices of compactification manifold can be thought of as being separated by domain walls. The Swampland Cobordism Conjecture \cite{McNamara:2019rup} formalises this picture by postulating that such domain walls have to exist. 

However, domain walls can be thin or thick; the latter can be described as a kink solution with fields gradually moving from one vacuum to another. When the thickness of the domain wall is much smaller than the short-distance cutoff of the effective field theory (EFT), we can regard the wall as thin. 

Consider for instance what happens in the absence of moduli stabilisation for the landscape of
4d supersymmetric (SUSY) $\mathcal{N}=2$ vacua from type II string theory on Calabi-Yau 3-folds.
According to Reid's conjecture \cite{Reid}, all Calabi-Yau spaces can be connected to each other through geometric transitions which can change the topology. From a 4d viewpoint, these transitions can be seen as motion in the space of (potentially massive) scalars from one vacuum to another. 

In this letter we continue on our previous work \cite{Shiu:2022oti} and argue for a similar picture in the presence of moduli stabilisation through fluxes. In other words: thin domain walls separating flux quanta can be seen as thick domain walls by integrating in certain open-string degrees of freedom. Hence, in the larger field space containing (possibly very heavy) open-string fields, we arrive at a picture in which different flux vacua are different local minima inside a single scalar potential.
In ref. \cite{Shiu:2022oti}, we demonstrated this for the original scale-separated flux vacua of type IIA orientifold compactifications of refs. \cite{DeWolfe:2005uu, Camara:2005dc, Derendinger:2004jn, Villadoro:2005cu}. This was partially motivated by an ongoing debate on the consistency of these vacua as string-theoretic backgrounds. The open-string fields in this case come from unstable D4-branes wrapping trivial 1-cycles inside non-trivial 2-cycles Poincar\'e dual to the four-form fluxes of the would-be solution. The positions of these branes correspond to 4d scalars whose scalar potential variations are smaller than the size of the actual vacuum energy.

The present work aims to generalise this picture to any setup. Let us consider some flux vacuum characterised by a specific $n$-dimensional internal manifold and a set of integers describing flux quanta over various submanifolds. Using Hodge duality, one can describe any flux as some internal $k$-form $F_k$, with $k\leq n$, threading compact dimensions only. If this form is for instance an RR-flux, then a D$(8-k)$-brane wrapping the dual $(n-k)$-cycle would appear as a domain wall in the non-compact $(10-n)$-dimensional spacetime separating vacua with RR-flux quanta differing by the D$(8-k)$-brane charge. A universal way to think of this domain wall as a thick wall is by regarding the D$(8-k)$-brane as a tachyon kink solution of an unstable
$\tilde{\mathrm{D}}(9-k)$-brane \cite{Sen:1999mg}. In other words, one integrates in open-string tachyons and, in the enlarged field space of compactification moduli and tachyons, the different vacua reside in one single scalar potential. 

Besides the general conceptual picture, another motivation for our work   originates in the so-called Anti-de Sitter (AdS) Distance Conjecture \cite{Lust:2019zwm}, which states that, for any $d$-dimensional
AdS flux vacuum with cosmological constant $\Lambda_{\mathrm{AdS}}$, there exists a tower of states with a mass scale $m$ that behaves as
\begin{equation}
    \dfrac{m}{m_{\p, d}} \overset{\Lambda_{\mathrm{AdS}} \sim 0}{\sim} \ab m_{\p, d}^{-2} \Lambda_{\mathrm{AdS}}\ab^{\alpha},
\end{equation}
for a positive constant $\alpha$, where $m_{\p, d}$ is the Planck mass. The strong form of this conjecture states that $\alpha=1/2$ for SUSY vacua. Whereas the general conjecture is supported by every controlled vacuum solution ever constructed, the strong form is more controversial and even more far-reaching as it would imply that SUSY AdS vacua do not admit a separation of scales and cannot be regarded as vacua inside a lower-dimensional EFT as the tower of particles does not decouple. Since SUSY AdS vacua are known to have CFT duals,
the implications for holography are equally far-reaching. 

The argument for the general form of the conjecture relies on an extension of the ordinary Distance Conjecture \cite{Ooguri:2006in}, which quantifies how a tower of states becomes light when the geodesic distance travelled in moduli space becomes large in Planck units. The generalisation that leads to the AdS Distance Conjecture of ref. \cite{Lust:2019zwm} relied on distances in metric space instead of (scalar) moduli space. Unfortunately, there is no independent support for the interesting suggestion that the Distance Conjecture can be applied to fields of all spin.

In light of this, it is instructive to connect flux vacua through trajectories in scalar field space since then one can use the kinetic term of the scalar fields to compute the distance between different vacua. This was done in ref. \cite{Shiu:2022oti} for the scale-separated IIA flux vacua of refs. \cite{DeWolfe:2005uu, Camara:2005dc, Derendinger:2004jn, Villadoro:2005cu}, and it was found that the Distance Conjecture is obeyed
by the scalars that provide the interpolation of flux quanta.
We refer to refs.~\cite{Farakos:2023nms, Tringas:2023vzn,Farakos:2023wps} for extensions of these observations to other vacua and to refs.~\cite{Li:2023gtt, Basile:2023rvm} for alternative measures of 
distances between vacua.

In this paper, using unstable $\tilde{\mathrm{D}}p$-branes, we can demonstrate in general circumstances that flux vacua can be seen as critical points in a single potential and how this realises the exponential dependence on the distance of the tower mass-scale. It furthermore verifies the suggestion \cite{Ooguri:2006in} that the field space metric becomes hyperbolic in the large-distance limit. However, since the field trajectories involve massive fields, strictly speaking, we cannot see this as a manifestation of the original Distance Conjecture for moduli spaces but its stronger and more speculative extension to fields which are lifted in a scalar potential.

\section{Non-BPS branes as domain walls}
\nointerlineskip
In addition to the RR-charged stable BPS D-branes, type II theories also feature non-BPS $\tilde{\mathrm{D}}$-branes, which are unstable D$p$-branes for $p$ odd/even in type IIA/IIB theories; for a review, see e.g. refs. \cite{Sen:1999mg, Harvey:2000qu, Sen:2004nf}. As a manifestation of the instability of such non-BPS branes, the spectrum of open strings ending on the latter is non-supersymmetric and contains a tachyon along with a gauge field.

At small string coupling, unstable non-BPS $\tilde{\mathrm{D}}p$-branes are described by a worldvolume action that resembles in many aspects the action of stable BPS D-branes. First, they have a string-frame Dirac-Born-Infeld term \cite{Sen:2004nf}
\begin{align}
    S_{\mathrm{DBI}}^{\tilde{\mathrm{D}}p} = - \tilde{\mu}_p \int_{\Sigma_{1,p}} \de^{1,p} \xi \; \e^{-\Phi} \, V(T) \, \sqrt{- \mathrm{det} \, \bigl[ \gamma_{\alpha \beta} \bigr]},
\end{align}
where $\smash{\gamma_{\alpha \beta} = G_{\alpha \beta} + \der_\alpha T \der_\beta T + (B + 2 \pi \ell_s F)_{\alpha \beta}}$. Here, $\Phi$ is the 10-dimensional dilaton, while $G_{\alpha \beta}$ and $B_{\alpha \beta}$ are the metric and Kalb-Ramond field. Furthermore, $F = \de A$ is the field-strength of the open-string gauge field $A_\alpha$ and $T$ is the open-string tachyon, while $\tilde{\mu}_p = \sqrt{2} \, \mu_p = \sqrt{2}\, 2\pi/(2\pi \ell_s)^{p+1}$ is the tension of the brane. Finally, $V(T)$ is the function
\begin{align}
\label{eq:tachyon_pot}
    V(T) = \dfrac{1}{\displaystyle \mathrm{cosh} \, \biggl(\dfrac{T}{\sqrt{2} \ell_s}\biggr)},
\end{align}
which represents the tachyon potential. Second, the Wess-Zumino term, in poly-form notation, is \cite{Billo:1999tv, Sen:2004nf}
\begin{equation}
    S_{\mathrm{WZ}}^{\tilde{\mathrm{D}}p} = \tilde{\mu}_p\int_{\Sigma_{p+1}} C \wedge \mathrm{tr} \, \bigl[ W(T) \, \de T \wedge \e^{B + 2 \pi \ell_s F} \bigr]\,,
\end{equation}
where $W(T)$ is a function that behaves like 
\begin{equation}
    W(T) \overset{T \sim \pm \infty}{\sim} \dfrac{1}{2} \, \e^{\mp \frac{T}{\sqrt{2} \ell_s}}.
\end{equation}
A kink solution of the theory interpolates between the vacua at $T = \pm \infty$; such a kink solution, once plugged back into the action, represents a lower-dimensional stable D$(p-1)$-brane since $\smash{\tilde{\mu}_p \int_{-\infty}^\infty \de \xi \, V(T(\xi)) = \mu_{p-1}}$, and in the WZ-term we have $\tilde{\mu}_p\int_{-\infty}^\infty W(T)  \,\de T = \mu_{p-1}$.

Given this quick summary, let us consider a type II compactification of the form $\mathcal{M}_d \times \mathrm{X}_{n}$, with $d+n=10$, and imagine that there is some (unbounded) RR-flux $F_k$ winding a $k$-cycle $\Sigma_k$ inside the space $\mathrm{X}_{n}$.
Now imagine a spacetime-filling non-BPS $\tilde{\mathrm{D}}(9-k)$-brane wrapping the $(n-k)$-cycle dual to $\Sigma_k$. Then, changing the tachyonic worldvolume scalar field from one tachyon vacuum to another induces a change in $F_k$ flux by one unit.

This is motivated due to the fact that the non-BPS $\tilde{\mathrm{D}}(9-k)$-brane sources the Bianchi identity for $F_k$, being
\begin{equation}
    \dd F_k = 2 \kappa_{10}^2 \, \tilde{\mu}_p \, W(T) \, \dd T \wedge \delta(\Sigma_k),
\end{equation}
where we ignored the term $H_3 \wedge F_{k-2}$ as such a term is
absent in the examples treated in this paper.  The Bianchi identity can be integrated to become
\begin{equation}
    \dfrac{1}{(2\pi \ell_s)^{k-1}} F_k =  U(T) \, \delta(\Sigma_k) + N \epsilon_k,    
\end{equation}
where we defined $U(T) = \sqrt{2} \int_{-\infty}^{T} W(S) \dd S / (2\pi \ell_s)$ and $N$ is the constant background flux on the background-volume form $\epsilon_k$. By integrating over the $k$-cycle $\Sigma_k$, we obtain
\begin{equation}
    \dfrac{1}{(2\pi \ell_s)^{k-1}} \int_{\Sigma_k} F_k = U(T) + N. 
\end{equation}
Since $U(T = - \infty)=0$, the quantised flux is just $N$, but when $U(T=+\infty)=1$, there are $N+1$ units of flux.

One might worry that the flux vacuum we end up with is not a genuine vacuum due to the presence of the non-BPS brane. However, following Sen's conjecture \cite{Sen:1998sm}, a non-BPS brane with the tachyon condensed in its tachyon vacuum corresponds to not having any brane at all. This means that the non-BPS brane only exists during the transition, and one truly starts and ends with a flux vacuum without extra local objects. 

For all practical purposes, one can think of the non-BPS $\tilde{\mathrm{D}}(9-k)$-brane as a dimensional extension of a D$(8-k)$-brane domain wall in spacetime, wrapping the same $(n-k)$-cycle $\Sigma_{n-k}$. Such a brane is the usual domain wall that is responsible for interpolations between $F_k$-flux vacua.

As an example, let us have a look at the Freund-Rubin solution $\mathrm{AdS}_5 \times \mathrm{S}^5$ in type IIB supergravity. The solution is parameterised by the integer $N$, measuring the quantised self-dual $F_5$-flux. Additionally, there is the massless 10d dilaton. The AdS scale $L_{\mathrm{AdS}}$ and the radius $L$ of the $\mathrm{S}^5$ are related and both depend on the flux as $L_{\mathrm{AdS}} \sim N^{1/4} \ell_s$ and $L\sim N^{1/4} \ell_s$. According to our proposal, the tachyon of a spacetime filling non-BPS $\tilde{\mathrm{D}}4$-brane does the trick of controlling the $F_5$-flux on the dual 5-cycle, i.e. the $\mathrm{S}^5$.  

\section{Field-space metrics and distances}
\label{sec: field-space metrics}
As we argued above, moving in tachyon space allows us to change the RR-fluxes in the compactification. However, changing the flux alone does not guarantee one to end up in a new flux vacuum. Indeed, the scalar fields in the compactification need to adjust accordingly. For example, in $\mathrm{AdS}_5 \times \mathrm{S}^5$, the volume field of the $\mathrm{S}^5$, or its radius $L \sim N^{1/4} \ell_s$, also shifts when the flux is changed.
The goal of this section is to identify what the scalar field space, enlarged with the tachyons, looks like. We argue that the field space takes a hyperbolic structure for unbounded fluxes.
It is then natural to contemplate distances between the different flux vacua. However, there is a potential involved, 
hence the field space is not a moduli space. For the latter, a well-defined notion of distance exists as the geodesic distance in the moduli space: for a theory of scalar fields $\phi^a$, when moving along a trajectory in the moduli space, parametrised by $s$, the dimensionless field-space distance is
\begin{equation}
    \Delta = \int_{0}^{1} \dd s \; \kappa_d \sqrt{g_{a b} \, \dfrac{\de \phi^a}{\de s} \dfrac{\de \phi^b}{\de s}},
\end{equation}
where $g_{a b}$ is the field-space metric and $\kappa_d$ is the $d$-dimensional gravitational coupling. This definition has been extensively used in the context of the Swampland Distance Conjecture. It is less clear what a good definition of distance is in field spaces with potentials (see ref. \cite{Klaewer:2016kiy} where it was argued that the same formulas should apply; see also refs. \cite{Li:2023gtt, Basile:2023rvm} for alternatives).

We want to emphasise that we keep an open mind to what the precise definition of field distances should be. Nevertheless, if we take the geodesic distance in the field space, we find agreement with the Swampland Distance Conjecture, due to the hyperbolic structure
of the field-space geometry.

An expansion of the DBI-action of a non-BPS $\tilde{\mathrm{D}}(9-k)$-brane shows that the field-space metric\footnote{In this note, we follow the standard convention in which the field-space metric $g_{a b}$ of a set of scalar fields $\phi^a$ can be read off from the kinetic action
\begin{align*}
    S = - \int_{M_{1,d-1}} \de^{1,d-1} x \sqrt{-\tilde{g}_{1,d-1}} \; \dfrac{1}{2} \, g_{a b}(\phi) \, \tilde{g}^{\mu \nu} \der_\mu \phi^a \der_\nu \phi^b.
\end{align*}
} for the tachyon $T$ reads
\begin{align}
    g_{TT} = \dfrac{2 \pi \sqrt{2}}{(2 \pi \ell_s)^{d}} \, \dfrac{\mathcal{V}_{n-k}}{\sqrt{\mathcal{V}_n}} \, \e^{\Phi_d} \, V(T).
\end{align}
Here, $\Phi_d$ is the $d$-dimensional dilaton and $\mathcal{V}_{n-k}$ and $\mathcal{V}_{n}$ are the dimensionless string-frame volumes of the wrapped cycle and of the total internal space, respectively. Below, we redefine the tachyon as $\smash{X \equiv \gamma \int \dd T \, \sqrt{V(T)}}$, where the factor $\gamma$ absorbs all constant factors and the appropriate dimensionful terms, except for a factor of $\sqrt{2}$ for convenience. If, for instance, the internal volume is factorised such that two string-frame radions $\sigma_1$ and $\sigma_2$ control such volumes as $\smash{\mathcal{V}_{n-k} = \e^{(n-k) \sigma_1}}$ and $\smash{\mathcal{V}_n = \e^{(n-k) \sigma_1 + k \sigma_2}}$,\footnote{The total volume can be controlled by several radions, but here a simple factorised structure of the internal space has been considered for simplicity; in the explicit examples, we will be precise.} the tachyon metric then becomes
\begin{align}
    g_{XX} = 2 \, \e^{\frac{\sqrt{d-2}}{2} \, \tdelta_d + \frac{\sqrt{n-k}}{2} \, \tsigma_1 - \frac{\sqrt{k}}{2} \, \tsigma_2},
\end{align}
where we also canonically normalised the closed-string scalars.\footnote{A review of how to canonically normalise the dilaton and the radions is in appendix \ref{app: canonical normalization of closed-string scalars}. For brevity, in the main text, we will drop the factors of $\kappa_d$.} Now, the tachyon vacua at $\smash{T = \pm \infty}$ correspond to vacua for $X$ at finite values, which after fixing the integration constant in the definition of $X$, can be taken at $\smash{X= 0, X_0}$. Furthermore, we can imagine extending the range of $X$ from this interval to the real line $\smash{\mathbb{R}}$, where the points $\smash{X = N X_0}$, for $\smash{N = 0, 1, \dots}$, correspond to vacua where the associated flux is $\smash{N}$.

In the case of a single unbounded flux, the field space clearly contains a hyperbolic factor $\mathbb{H}^2$ since there is one open-string tachyon which is of course coupled to a single linear combination of closed-string fields (dilaton and radions). This is actually a more general structure. If there are $m$ fluxes that are unbounded, and which are not related to each other by tadpole conditions, then the field space contains a hyperbolic factor $\smash{(\mathbb{H}^2)^m}$: this is because the kink solutions for each of the tachyons controlling one of the $m$ flux quanta would allow one to make a jump in this flux without affecting the others. Therefore, the tachyon kinetic couplings must be orthogonal. In other words, the field-space metric is always of the form
\begin{equation}
    \dd s_{\text{field}}^2 = 2 \sum_{r=1}^m \e^{\sum_a \alpha_{r a} \varphi^a} \, (\dd X_r)^2 + \sum_a (\dd \varphi^a)^2,
\end{equation}
where $\smash{\varphi^a=(\tdelta_d, (\tsigma_i)_i)}$ represents the canonically-normalised closed-string moduli, and is such that
\begin{equation}
    \sum_a \alpha_{r a} \alpha_s{}^a = \delta_{rs}.
\end{equation}

In the remainder of this section, we verify this behaviour for some examples.
We illustrate it here for the simplest scale-separated vacua of massive type IIA string theory and for $\mathrm{AdS}_5 \times \mathrm{S}^5$ vacua of type IIB string theory. We also compute the scalar field distances in the large flux regime. In app. \ref{App:distances}, we explain how these distances can be calculated, and we show that they are always logarithmic in the flux quantum $N$.
Additionally, we performed an analogous analysis for the scale-separated vacua in massless type IIA of ref. \cite{Cribiori:2021djm} and for $\mathrm{AdS}_3 \times \mathrm{S}^3 \times \mathbb{T}^4$ vacua of type IIB in app. \ref{app:other_vacua}. In all of these examples, we only consider the scalars whose vacuum expectation values shift under flux changes.

\subsubsection*{$\mathrm{AdS}_4 \times \mathrm{CY}_3$ vacua in massive type IIA}
Among our main examples of interest are the flux solutions on a toroidal orientifold (or more generally on Calabi-Yau orientifolds) of massive type IIA string theory found in refs. \cite{DeWolfe:2005uu, Camara:2005dc, Derendinger:2004jn, Villadoro:2005cu}. There are three unbounded $F_4$-fluxes on three different 4-cycles. Following the prescription outlined above, we can wrap non-BPS $\tilde{\mathrm{D}}5$-branes on the 2-cycles dual to the four-cycles. Therefore, we have three different tachyons $X_i$ in the problem, with $i$ corresponding to the $i$-th 2-cycle that the brane is wrapping, whose string-frame radions are denoted as $\tsigma_i$. The field-space metric is
\begin{equation}
    \dd s_{\text{field}}^2 = 2 \sum_i \e^{\frac{\tdelta_4 + \sum_j \epsilon_{ij} \tsigma_j}{\sqrt{2}}} \dd X_i^2 + \dd \tdelta_4^2 + \sum_i \dd \tsigma_i^2,
\end{equation}
where $\epsilon_{ij} = - 1 + 2 \delta_{ij}$,
and by the $\mathrm{SO(4)}$-rotation
\begin{subequations}
    \begin{align}
        P_i & \equiv \frac{1}{2} \, \bigl( \tdelta_4 + \sum_j \epsilon_{ij} \tsigma^j \bigr), \\
        Q & \equiv \frac{1}{2} \, \bigr( \tdelta_4 + \tsigma_1 + \tsigma_2 + \tsigma_3 \bigr)\,,
\end{align}
\end{subequations}
and after defining $P_i = - \sqrt{2} \, \mathrm{log} Z_i$, it becomes
\begin{equation}
    \dd s_{\text{field}}^2 = 2\sum_i \frac{\dd X_i^2 + \dd Z_i^2}{Z_i^2} + \dd Q^2.
\end{equation}
As expected, the field space has the form $(\mathbb{H}^2)^3\times \mathbb{R}$. Therefore, the geodesic distance can be calculated with the techniques illustrated in app. \ref{App:distances}.
As $\tilde{\sigma}_i \sim (\sqrt{2}/4) \, \log N$ and $\tilde{\Phi}_4 \sim -(3\sqrt{2}/2) \, \log N$, we find that $Z_i \sim N^{7/8}$ and $Q \sim (3/8) \, \log N$ and still $X_i \sim N$, such that the distance becomes $\Delta = \sqrt{63/8} \, \log N$. Notice that the distance is not only logarithmic in $N$, but also smaller than what was found in ref. \cite{Shiu:2022oti}.

This logarithmic dependence ensures that the Distance Conjecture is satisfied. Indeed, the Distance Conjecture applied to the Kaluza-Klein (KK) tower, taking the form $\smash{m_\mathrm{KK}(N)/m_{\mathrm{KK},0} = \e^{-\alpha \Delta}}$, agrees with the power-law scaling of $\smash{m_\mathrm{KK}(N) \sim N^{- 1/4} \ell_s}$. Here  $m_{\mathrm{KK},0}$ is the KK mass one starts from. Furthermore, we find that $\smash{\alpha = \sqrt{1/126} < 1/\sqrt{2} }$, violating the sharpened Distance Conjecture of ref. \cite{Etheredge:2022opl} proposed for exact moduli spaces.

\subsubsection*{$\mathrm{AdS}_5 \times \mathrm{S}^5$ in type IIB}
In the $\mathrm{AdS}_5 \times \mathrm{S}^5$ solutions of type IIB strings, it will be a non-BPS $\tilde{\mathrm{D}}4$-brane that has the right couplings for changing the $F_5$ flux. This $\tilde{\mathrm{D}}4$-brane fills spacetime and has no cycle to wrap in the internal space. The relevant field space is made up of the 5d volume, the dilaton and the tachyon and it reads
\begin{equation}
    \dd s_{\text{field}}^2 = 2 \, \e^{\frac{\sqrt{3}}{2} \tdelta_5 - \frac{\sqrt{5}}{2} \tsigma} \dd X^2 + \dd \tdelta_5^2+ \dd \tsigma^2.
\end{equation}
After the $\mathrm{SO}(2)$-rotation
\begin{equation}
    \tvec{P}{Q} \equiv \dfrac{1}{2\sqrt{2}} \matr{\sqrt{3}}{-\sqrt{5}}{\sqrt{5}}{\sqrt{3}} \tvec{\tdelta_5}{\tsigma},
\end{equation}
and redefining $P = - \sqrt{2} \, \mathrm{log} \, Z$, we find
\begin{equation}
    \dd s_{\text{field}}^2 = 2 \, \dfrac{\dd X^2 + \dd Z^2}{Z^2} + \dd Q^2\,.
\end{equation}
Again, the field space is the product of the hyperbolic plane and the real line, i.e. $\mathbb{H}^2\times \mathbb{R}$. From the proper scalings, $\tilde{\sigma} \sim (\sqrt{5}/4) \, \log N$ and $\tilde{\Phi}_5 \sim -(5/4\sqrt{3}) \, \log N$ one finds that $Z \sim N^{5/8}$ and $Q \sim (1/4) \, \log N$ and $X \sim N$, from which it follows that the distance is $\Delta = \sqrt{23/6} \, \log N$. Here the logarithmic behaviour also ensures that the Distance Conjecture is satisfied. Using that $\smash{m_\mathrm{KK} \sim N^{- 1/4} \ell_s}$, we find that $\smash{\alpha = \sqrt{3/184}}$ also violating the sharpened Distance Conjecture of ref. \cite{Etheredge:2022opl}.

\section{Discussion}
We have argued on general grounds, and
shown
with concrete examples, that vacua that differ in flux quanta can be seen as different critical points of a single-valued scalar potential, extending ideas of ref. \cite{Shiu:2022oti}. This allows us to define, in a natural way, the distance between flux vacua as a distance travelled in scalar field space. This can be done by integrating in open string fields. In this paper, we took them to be tachyon fields of unstable $\tilde{\mathrm{D}}p$-branes, but this choice is not unique. In our previous work \cite{Shiu:2022oti}, we took those open-string fields to be the positions of $\mathrm{D}p$-branes wrapping cycles trivial in homology.
A benefit of using tachyon fields of unstable $\tilde{\mathrm{D}}p$-branes is that one can argue more model-independently how flux vacua arise as critical points in a universal single-valued scalar potential. 

We suggest that requiring interpolations between vacua to be continuous postdicts the existence of these non-BPS branes, in the spirit of the Swampland Cobordism Conjecture \cite{McNamara:2019rup}.
However, string theory has more to offer than vacua where the moduli are stabilised with RR-fluxes alone. There exist scenarios where unbounded NSNS-fluxes do stabilise the moduli, such as e.g. in the non-SUSY vacua of ref. \cite{Baykara:2022cwj}, and M-theory has vacuum solutions with $G_7$- and $G_4$-fluxes. If a continuous version of the Cobordism Conjecture is to be taken seriously, then our proposal implies the existence of extended \emph{non-BPS} objects with NS5-charge and non-BPS objects with $\mathrm{M}2$- or $\mathrm{M}5$-charges in M-theory. The idea is then again that BPS thin domain walls are described as thick domain walls that lift to space-filling non-BPS branes.

For the examples that we have studied, this formalism confirms that the field space connecting asymptotic vacua in the large-flux regime is hyperbolic and the distances are logarithmic in the flux quantum $N$. Since the KK-masses scale with a negative power of $N$ this agrees with the ordinary Swampland Distance Conjecture \cite{Ooguri:2006in}, stating that the mass-scale of the (KK-)tower is exponential in the distance, and not just the AdS Distance Conjecture \cite{Lust:2019zwm}. Note that we do not touch upon the \emph{strong} AdS Distance Conjecture of ref. \cite{Lust:2019zwm}, which rules out supersymmetric scale-separated vacua.

One could argue that there is not as much evidence for a Distance Conjecture applied to scalar fields that are not pure moduli but are lifted by a scalar potential instead. In such cases, we tend to use the Distance Conjecture only when the potential is exponentially suppressed in asymptotic regimes.  Below we argue why, despite the naive high mass scale of the tachyon fields, there is a notion in which they behave as pure moduli in the large-$N$ limit, which could be the explanation of why the ordinary Distance Conjecture is obeyed.

\begin{figure}[ht]
    \centering
    
    \begin{tikzpicture}[xscale=0.65,yscale=1.00]

    \draw[gray,densely dotted] (1,-6.3) -- (1,0);
    \draw[gray,densely dotted] (3,-5.2) -- (3,0);
    \draw[gray,densely dotted] (5,-4.1) -- (5,0);
    \draw[gray,densely dotted] (7,-3) -- (7,0);
    \draw[gray,densely dotted] (9,-2) -- (9,0);
    \draw[gray,densely dotted] (11,-1.1) -- (11,0);
    
    \draw[->] (-0.5,0) -- (11.5,0) node[below]{$X$};
    \draw[->] (0,-5.5) -- (0,1) node[above]{$V_\text{total}$};

    \draw[thick,color=orange] (0.5,-5.8) arc (180:360:0.5);

    \draw[thick,color=orange] (1.5,-5.8) -- (1.5,-4.2) to[out=90,in=90] (2.7,-4.1) --(2.7,-4.9);

    \draw[thick,color=orange] (2.7,-4.9) arc (180:360:0.3);

    \draw[thick,color=orange] (3.3,-4.9) -- (3.3,-3.5) to[out=90,in=90] (4.8,-3.3) -- (4.8,-3.9);

    \draw[thick,color=orange] (4.8,-3.9) arc (180:360:0.2);

    \draw[thick,color=orange] (5.2,-3.9) -- (5.2,-2.8) to[out=90,in=90] (6.9,-2.5) -- (6.9,-2.9);

    \draw[thick,color=orange] (6.9,-2.9) arc (180:360:0.1);

    \draw[thick,color=orange] (7.1,-2.9) -- (7.1,-2.3) to[out=90,in=90] (8.95,-1.75) -- (8.95,-1.95);    

    \draw[thick,color=orange] (8.95,-1.95) arc (180:360:0.05);

    \draw[thick,color=orange] (9.05,-1.95) -- (9.05,-1.85) to[out=90,in=90] (10.975,-1.175);

    \draw[thick,color=orange] (10.975,-1.175) arc (180:360:0.025);

    \draw[thick,color=orange] (11.025,-1.175) to[out=90,in=260] (11.025,-0.75);

    \draw[->] (9,0.25) -- (10,0.25) node[above,pos=0.5]{$N \to \infty$};

    \end{tikzpicture}
    
    \caption{The figure shows the main qualitative features of the potentials that we consider in this paper. The different local minima correspond to AdS vacua corresponding to different flux units $N$, while the wiggles represent the contribution to the total potential coming from the potential along the tachyonic direction $X$. As $N$ grows, the tachyon potential that lets the theory hop between vacua with different flux units has three main features: the local minima are equidistant, the convexity of the minima increases (with respect to the KK-scale), and the height of the wiggles decreases.}
    \label{fig: wiggled potential}
\end{figure}
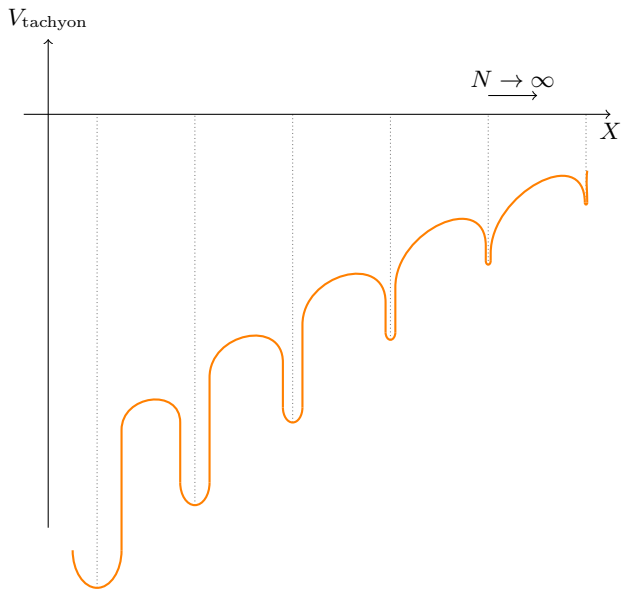

The scalar potential of the tachyon in the $d$-dimensional Einstein frame scales as
\begin{equation}
    V_\text{tachyon} \sim \dfrac{2 \pi \sqrt{2}}{(2 \pi \ell_s)^d} \frac{\mathcal{V}_{n-k}}{\sqrt{\mathcal{V}_n}} \, \e^{\frac{d+2}{2 \sqrt{d-2}} \tdelta_d} \, V(T).
\end{equation}
Therefore, because $\smash{\der^2 V_\text{tachyon}/\der T^2 \sim V_\text{tachyon}/(2 \ell_s^2)}$, we can estimate the scaling of the tachyon mass in $d$-dimensional Planck units as
\begin{equation}
    m_\text{tachyon}^2 \sim \dfrac{1}{2} \, g_{TT}^{-1} \dfrac{\der^2}{\der T^2} V_\text{tachyon} \sim \ell_s^{-2} \, \e^{\frac{2 \, \tdelta_d}{\sqrt{d-2}}},
\end{equation}
as we should have expected since the mass scale of the tachyon is the string scale. On the other hand, the KK-mass in $d$-dimensional Planck units scales as
\begin{equation}
    m_\mathrm{KK}^2 \sim {L_\mathrm{KK}^{-2}} \, \e^{\frac{2 \, \tdelta_d}{\sqrt{d-2}}},
\end{equation}
where $L_\mathrm{KK}$ is the radius in string units of the largest cycle in the compact manifold. Since we require large volumes, we see that the effective tachyon mass is higher than the KK-scale. One might worry that this should invalidate the whole effective field theory, as the KK-scale provides an upper bound for the cut-off scale of the theory. However, we will argue that this is not necessarily true. Indeed, we should also compare the tachyon energy with the energy coming from the flux potential. How this scales with the flux parameter $N$ depends on the compactification.
For instance, for the simple scale-separated flux vacua in type IIA string theory \cite{DeWolfe:2005uu, Camara:2005dc, Derendinger:2004jn, Villadoro:2005cu}, the tachyon potential is parametrically smaller than the flux potential, being
\begin{equation}
    \frac{V_\text{tachyon}}{V_{\mathrm{AdS}_4 \times \mathrm{CY}_3}} \sim N^{-1/4}.
\end{equation}
For the $\mathrm{AdS}_5 \times \mathrm{S}^5$ vacua of type IIB strings, we have a similar pattern, being
\begin{equation}
    \frac{V_\text{tachyon}}{V_{\mathrm{AdS}_5 \times \mathrm{S}^5}} \sim N^{-3/4}.
\end{equation}
This means that, in the tachyon direction $X$, both examples have tachyon vacua with three very specific features:
\begin{enumerate*}[label=(\roman*)]
    \item they are equidistant, when we are considering just the $X$-direction while keeping the $Z$-field constant;
    \item they are separated by hills in the potential that become smaller in size;
    \item they become sharper with respect to the KK-scale, as $N$ becomes larger.\footnote{This specific wiggly behaviour does not happen for all examples, as shown in app. \ref{app:other_vacua}.}
\end{enumerate*}
This is illustrated qualitatively in fig. \ref{fig: wiggled potential}.

In this particular sense the tachyon field behaves as a modulus at large $N$: the wiggles (i.e. energy differences between the vacua) become arbitrarily small in AdS units. Hence, the tachyon field can be given a kinetic energy well below the AdS-scale so that it will fly through the potential as if the wiggles were absent. This is not different from our usual notion of asymptotic field variations in compactifications and the usual Distance Conjecture should apply. Yet, one could object that the second derivative of the canonically normalised tachyon field remains of order of the string scale in every vacuum. This is small in Planck units since the string scale is suppressed by inverse powers of $N$ with respect to the Planck scale. Yet, this is large in KK units, which is our EFT cut-off. Since the tachyon is not part of the $d$-dimensional supergravity multiplet we consider it possible that its mass renormalisation is strong enough for it to obtain a mass below the cut-off, but this is merely a suggestion at this point.

Finally, our arguments should apply beyond vacua with unbounded fluxes. One could for instance consider the type IIB Landscape of 4d vacua from compactifications on conformal Calabi-Yau three-folds. The complex structure moduli are stabilised by three-form fluxes which are bounded by tadpole constraints \cite{Dasgupta:1999ss, Giddings:2001yu}. One could interpolate between such flux vacua by using non-BPS branes, or by space-filling 5-branes as in KPV \cite{Kachru:2002gs}.
It would be interesting to investigate how scalar field distances in this context relate to obtaining a small on-shell flux superpotential $W_0$ for KKLT constructions as in ref. \cite{Demirtas:2021ote}, but we leave this to further research.

\begin{acknowledgments}
\subsection*{Acknowledgments}
GS is supported in part by the DOE
grant DE-SC0017647. FT and TVR are supported by the FWO Odysseus grant GCD-D0989-G0F9516N. VVH is supported by the Olle Engkvists Stiftelse.
\end{acknowledgments}

\appendix

\section{Canonical normalization of closed-string scalars} \label{app: canonical normalization of closed-string scalars}

Let $\ell_s = \sqrt{\alpha'}$ be the string length. We can define the $d$-dimensional Einstein-frame metric $\smash{\tilde{g}_{\mu \nu}}$, which determines a line element expressed as $\smash{d \tilde{s}_{1,d-1}^2 = \tilde{g}_{\mu \nu} \, \de x^\mu \de x^\nu}$, by parameterising the string-frame metric, defined through the line element $d s_{1,9}^2 = G_{MN} \de x^M \de x^N$, as
\begin{equation} \label{d-dim. Einstein-frame metric (delta,sigma)}
    \begin{split}
        ds_{1,9}^2 = \e^{\frac{4 \Phi_d}{d-2}} \, \vevc \, \tilde{g}_{\mu \nu} \, \de x^\mu \de x^\nu + \e^{2 \sigma} \, \breve{g}_{m n} \, \de y^{m} \de y^{n},
    \end{split}
\end{equation}
where $\Phi_d$ is the $d$-dimensional dilaton and $\sigma$ is the string-frame radion.
Here, the internal metrics are assumed to be normalised in such a way that $\smash{\int_{\mathrm{K}_{n}} \de^n y \, \sqrt{\breve{g}_{n}} = (2 \pi \ell_s)^{n}}$, with $n=10-d$. An arbitrary constant $\vevc$ has also been inserted: fixing it as $\vevc = \e^{- 4 \langle \Phi_d \rangle / (d-2)}$ makes the $d$-dimensional metric components correspond in both frames, in the vacuum. Another possible choice is
\begin{equation} \label{d-dim. Einstein-frame metric (phi,omega)}
    ds_{1,9}^2 = \e^{\frac{\sdil}{2}} \, \biggl[ \dfrac{\vevc \, \tilde{g}_{\mu \nu} \, \de x^\mu \de x^\nu}{\e^{2 \omega \, \frac{10-d}{d-2}}} + \e^{2 \omega} \, \breve{g}_{m n} \, \de y^{m} \de y^{n} \biggr],
\end{equation}
which involves the shifted 10-dimensional dilaton $\sdil = \Phi - \mathrm{ln} \, g_s$, where $g_s = \e^{\langle \Phi \rangle}$ is the string-coupling vacuum expectation value, and the Einstein-frame radion $\omega$.
The $(\Phi_d, \sigma)$- and $(\sdil, \omega)$-bases can be related via the linear transformations $\smash{\Phi_d = [(d-2)/8] \, \sdil - [(10-d)/2] \, \omega}$ and $\smash{\sigma = \sdil/4 + \omega}$.
For an isotropic compactification, both bases are diagonal and the canonically normalised fields e.g. in the $(\Phi_d, \sigma)$-basis read
\begin{align*}
    \tdelta_d & = \dfrac{2}{\sqrt{d-2}} \, \dfrac{\Phi_d}{\kappa_d}, \\
    \tsigma & = \sqrt{10-d} \, \dfrac{\sigma}{\kappa_d}.
\end{align*}
Here, given the string-frame gravitational coupling $\smash{2 \kappa_{10}^2 = (2 \pi \ell_s)^8 / 2 \pi}$ (which is not proper, due to the extra dilaton factor), the reduced $d$-dimensional gravitational coupling is
\begin{equation} \label{d-dim. gravitational coupling}
    2 \kappa_d^2 = \dfrac{1}{2 \pi} \dfrac{g_s^2 \, \e^{2 \langle \Phi_d \rangle}}{(2 \pi \ell_s)^{2-d}} = \dfrac{g_s^2 \, (2 \pi \ell_s)^{d-2}}{2 \pi \, \e^{(10-d) \langle \omega \rangle}}.
\end{equation}
This is related to the $d$-dimensional Planck mass $\smash{m_{\p, d}}$ as $\smash{m_{\p, d} = \kappa_d^{2/{d-2}}}$.
Although the 10-dimensional dilaton and the Einstein-frame radions are more intuitive quantities to work with, in an anisotropic compactification there are multiple radions and they are kinetically-mixed in the $d$-dimensional Einstein frame; the $d$-dimensional dilaton and the string-frame radions are instead always diagonal. In sec. \ref{sec: field-space metrics}, for instance, we consider internal metrics decomposed as $\smash{ds_{n}^2 = \e^{2 \sigma_1} d \breve{s}_{n-k}^2 + \e^{2 \sigma_2} d \breve{s}_{k}^2}$. For the details of anisotropic compactifications, see e.g. ref. \cite[app. B]{Shiu:2023fhb}.

\section{Computing distances in hyperbolic space} \label{App:distances}
\begin{figure}[h]
    \centering
    \begin{tikzpicture}[dot/.style={draw,circle,minimum size=2mm,inner sep=0pt,outer sep=0pt,blue,fill=blue!50!cyan,solid}]
  
    \draw[->] (-3.5,0) -- (3.5,0) node[right] {$X$};
    \draw[->] (-3,-0.5) -- (-3,3.5) node[left] {$Z$};

    \draw[rotate=-10,cyan,densely dotted] (0,0) -- (180:3);
    \draw[rotate=-10,cyan,densely dotted] (0,0) -- (60:3) node[below right, pos=0.5,black]{$l$};
    \draw[rotate=-10,->,cyan,thin] (-0.6,0) arc (180:60:0.6); 
    
    \draw (-3,0) arc (180:0:3);
    \draw[rotate=-10,thick,color=blue!50!cyan] (-3,0) arc (180:60:3) node[dot,pos=0]{} node[above right,black,pos=0]{$s=0$} node[dot]{} node[above right,black]{$s=1$};
    
    \end{tikzpicture}
    \caption{A sketch of the geodesic path in hyperbolic plane.}
    \label{fig: semicircle}
\end{figure}
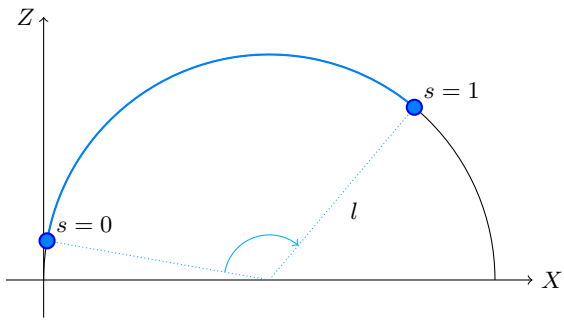

Suppose we have a hyperbolic field space of the form
\begin{equation}
    \dd s_{\text{field}}^2 = \frac{1}{Z^2} \bigl( \dd X^2 + \dd Z^2 \bigr).
\end{equation}
Let us suppose that we want to compute a distance in this field space along a path $(X(s),Z(s))$ where $X(0) \sim 0$, $X(1) \sim N^f$ and $Z(0) \sim 1$, $Z(1) \sim N^g$. Here, $N \gg 1$ is a large positive number, while $f$ and $g$ are constant powers one can specify for any given model of interest. Of course, $N$ has to be interpreted as a large flux at the end of the trajectory at $s=1$, whereas the beginning at $s=0$ would correspond to a low flux, say $N=1$.
The geodesics are arcs of semicircles, and the solution can be parameterised as
\begin{equation}
\label{eq:hyperbolic_geodesic}
    X(s) = l \, \tanh (d_1 s + d_2) +X_c , \quad Z(s) = l \, \mathrm{sech } (d_1 s + d_2).
\end{equation}
In particular, the geodesics are such that
\begin{equation}
    [X_c - X(s)]^2 + [Z(s)]^2 = l^2
\end{equation}
and the geodesic distance is simply
\begin{equation}
    \Delta = \int_0^1 \dd s \, \sqrt{\dfrac{1}{Z^2} \biggl[ \biggl(\dfrac{\dd X}{\dd s}\biggr)^2 + \biggl(\dfrac{\dd Z}{\dd s}\biggr)^2 \biggr]} = \ab d_1 \ab.
\end{equation}
A picture of the geodesic path is in fig. \ref{fig: semicircle}.

All the integration constants $l$, $d_1$, $d_2$ and $X_c$ are fixed by the boundary conditions, as we explain below. To do this, let us assume that one of the endpoints of the semicircle is (close to) the origin, and therefore $X_c \simeq l$, which is justified in the large $N$-limit, since we also have $l \gg 1$ as we show below. We can estimate the radius of the semicircle by noticing that
$\smash{[l-X(1)]^2 + [Z(1)]^2 = l^2}$ is solved for
\begin{equation}
    l = \frac{(X(1))^2 + (Z(1))^2}{2X(1)}
\end{equation}
Let us look now at the case $f>g$, for which $X$ becomes much larger than $Z$.  It means that the trajectory goes over the top of the semicircle. In this limit, the radius $l$ behaves as $X(1)$, and thus $l \sim N^f$.
Let us now look at the equation for $Z$, evaluated at $s=0$. We know that this should not scale with $N$ at leading order. We find
\begin{equation}
    Z(0) = l\, \text{sech}\, d_2 \sim
    N^f \:\text{sech}\, d_2,
\end{equation}
We see that $d_2$ must scale as $d_2 = \pm f \log N$ in the large-$N$ limit. Without loss of generality, we take the minus sign. We now evaluate $Z$ at $s=1$, and from the requirement that $Z(1)\sim N^g$, we deduce that there are two options, either
\begin{equation}
\label{eq:d1_options_1}
    d_1 = (2f - g) \log N \quad \text{or} \quad  d_1 = g \log N.
\end{equation}
It turns out that if $Z$ is to increase and decrease over the interval $s \in [0,1]$, one needs to choose the first option in eq. \eqref{eq:d1_options_1}. This is because the $\mathrm{sech}$-function requires an argument, here $d_1 s + d_2$, that is first negative and then positive with increasing $s$ for that behaviour.
Hence the geodesic distance in the large $N$-limit is
\begin{equation}
    \Delta = (2f - g) \log N.
\end{equation}
Suppose now that the opposite is true: $g>f$. This means that the $Z$-coordinate becomes much larger than $X$ in the large-$N$ limit. 
It means the endpoint is not yet reaching the top of the semicircle. 
The same formula for the radius $l$ applies, and we now see from this that
$l = N^{2g-f}$.
From $Z(0) \sim 1$, it then follows that
\begin{equation}
    d_2 \sim \pm (2g-f) \log N
\end{equation}
and again we take the minus sign without loss of generality.
Consequentially, from $Z(1) \sim N^g$ we get that either
\begin{equation} \label{eq:d1_options_2}
    d_1 = (3g-2f) \log N \quad \text{or} \quad
    d_1 = g \log N.
\end{equation}
This time, $Z$ is to increase but not decrease over the interval $s \in [0,1]$, for which the first option in eq. \eqref{eq:d1_options_2} needs to be taken, as the argument of the $\mathrm{sech}$-function needs to remain negative.
Hence the distance is now
\begin{equation}
    \Delta = g \log N.
\end{equation}

Suppose now that the field space is a direct product of hyperbolic planes and the real line, with metric
\begin{equation}
    \dd s_{\text{field}}^2 = \sum_i \frac{a_i}{Z_i^2}\left(\dd X_i^2 + \dd Z_i^2\right) + b\, \dd Q^2
\end{equation}
In that case, the solution to the geodesic equation is for all the $X_i$'s and $Z_i$'s the same as eq. \eqref{eq:hyperbolic_geodesic}, with associated integration constants $d_{1,i}$, $d_{2,i}$, $l_i$ and $X_{0,i}$. The integration constants can be found analogously to the above. Furthermore, $Q$ is solved as $Q(s) = d_3 s + d_4$. Finally, the distance becomes
\begin{equation}
    \Delta^2 = \sum_i a_i \, d_{1,i}^2 + b\, d_3^2\,.
\end{equation}

\section{Field spaces and distances for scale-separated vacua of massless type IIA strings and $\boldsymbol{\mathrm{AdS}_3 \times \mathrm{S}^3 \times \mathbb{T}^4}$ of type IIB strings}\label{app:other_vacua}
In this appendix, we discuss the field space and compute the distances for the scale-separated vacua of massless type IIA string theory and the $\mathrm{AdS}_3 \times \mathrm{S}^3 \times \mathbb{T}^4$ vacua of type IIB string theory.

\subsubsection*{Scale-separated $\mathrm{AdS}_4$ vacua in massless type IIA}
The scale-separated solutions of ref. \cite{Cribiori:2021djm} are set in massless type IIA string theory, compactified on a SU(3)-structure nilmanifold (the Iwasawa manifold). For generalisations, see ref. \cite{Carrasco:2023hta}. In the simplest example of ref. \cite{Cribiori:2021djm}, there are three unbounded fluxes: two $F_2$-fluxes and the $F_6$-flux. Following our prescription, we can use two non-BPS $\tilde{\mathrm{D}}7$-branes wrapping the two 4-cycles dual to the 2-cycles that the $F_2$-fluxes are winding to change them. Similarly, we can use a spacetime-filling non-BPS $\tilde{\mathrm{D}}3$-brane to change the $F_6$-flux. The field space metric is
\begin{align}
\begin{split}
    \dd s_{\text{field}}^2 = 2 \,\e^{\frac{\tdelta_4 - \tsigma_1 - \tsigma_2 - \tsigma_3}{\sqrt{2}}} \dd X_1^2 + 2 \, \e^{\frac{\tdelta_4 + \tsigma_1 + \tsigma_2 - \tsigma_3}{\sqrt{2}}} \dd X_2^2 \\
    + 2 \, \e^{\frac{\tdelta_4 + \tsigma_1 - \tsigma_2 + \tsigma_3}{\sqrt{2}}} \dd X_3^2 + \dd \tdelta_4^2 + \sum_i \dd \tsigma_i^2 & ,
\end{split}
\end{align}
where $X_1$ represents the D7 tachyon and $X_2$ and $X_3$ the D3 tachyons.
After an $\mathrm{SO}(4)$-rotation of the fields, i.e.
\begin{subequations}
    \begin{align}
        P_1 & \equiv - \dfrac{1}{2} \bigl( \tdelta_4 - \tsigma_1 - \tsigma_2 - \tsigma_3 \bigr), \\
        P_2 & \equiv - \dfrac{1}{2} \bigl( \tdelta_4 + \tsigma_1 + \tsigma_2 - \tsigma_3 \bigr), \\
        P_3 & \equiv - \dfrac{1}{2} \bigl( \tdelta_4 + \tsigma_1 - \tsigma_2 + \tsigma_3 \bigr), \\
        Q & \equiv - \dfrac{1}{2} \bigl( \tdelta_4 - \tsigma_1 + \tsigma_2 + \tsigma_3 \bigr),
\end{align}
\end{subequations}
and defining $P_i =- \sqrt{2} \, \mathrm{log} Z_i$, the field space metric becomes
\begin{equation}
    \dd s_{\text{field}}^2 = 2\sum_i \frac{\dd X_i^2 + \dd Z_i^2}{Z_i^2} +  \dd Q^2,
\end{equation}

which means that the field space is again of the form $(\mathbb{H}^2)^3\times \mathbb{R}$. 
According to ref. \cite{Cribiori:2021djm}, the scalars scale as $\tilde{\Phi}_4 \sim -[\sqrt{2}(a+b+c)/4] \, \log N$, $\tilde{\sigma}_1 \sim [\sqrt{2}(a-b-c)/4] \, \log N$, $\tilde{\sigma}_2 \sim [\sqrt{2}(a-b+c)/4] \, \log N$ and $\tilde{\sigma}_3 \sim [\sqrt{2}(a+b-c)/4] \, \log N$. Consequentially, $Z_1 \sim N^{(5a + b+ c)/8}$, $Z_2 \sim N^{(a + 5b+ c)/8}$, $Z_3 \sim N^{(a + b+ 5c)/8}$ and $Q \sim [(a + b+ c)/8] \, \log N$. The tachyons, playing the role of the fluxes, have to scale as $X_1 \sim N^a$, $X_2 \sim N^b$ and $X_3 \sim N^c$.
The expression for the distance now depends on the relative size of $a$, $b$ and $c$. They all need to be positive. Without loss of generality, we take $c \geq b$ and it follows from growing volumes that $a > c \geq b$ and also $a>2b$.
From that, it can be seen that $X_1$ always scales higher than $Z_1$, whereas $X_2$ always scales lower than $Z_2$. If $a+b > 3c$ , then $X_3$ scales lower than $Z_3$ too, and the distance becomes
\begin{equation}
    \Delta_1 = \frac{ \sqrt{31 a^2-2 a (b+c)+7 b^2+ 6 b c+7
   c^2}}{2 \sqrt{2}} \, \log N.
\end{equation}
In the other case, when $a+b<3c$, we find 
\begin{equation}
    \Delta_2 = \frac{ \sqrt{31 a^2-2 a (b+5c)+7 b^2-2 b c+31
   c^2}}{2 \sqrt{2}} \, \log N.
\end{equation}
These expressions are compatible with each other when $a+b=3c$.

Looking at the energy scales, whether the tachyon potentials are lower than the flux potential depends on the relative size of the fluxes. Indeed, we have
\begin{equation}
    \frac{V_{X_i}}{V_{\text{AdS}_4\times \text{Iw}_6}} \sim N^{(a + b + c)/4 - a\delta_{i1} - b \delta_{i2}-c \delta_{i3}}
\end{equation}
With constraints on the scaling exponents $a$, $b$ and $c$ discussed above, we notice that the potential for the D3-tachyon $X_1$ is always lower than the flux potential, whereas the D7-tachyon $X_2$ always scales higher. For the D7-tachyon $X_3$ it depends: if $a+b>3c$, its potential scales lower than the flux potential and higher otherwise. This is reminiscent of the distance computations above, where we needed to investigate when the $X_i$ scale higher or lower than the $Z_i$. 

\subsubsection*{$\mathrm{AdS}_3 \times \mathrm{S}^3 \times \mathbb{T}^4$}
We can also take a look at type IIB string theory compactified to AdS$_3 \times \mathrm{S}^3 \times \mathbb{T}^4$ with an $F_3$-flux along the $\mathrm{S}^3$ and an $F_7$-flux along the whole compact space. Here the relevant scalar fields are the 3d dilaton $\tilde \Phi_3$, and the volume moduli of the $\mathrm{S}^3$ and the $\mathbb{T}^4$, which we call $\tilde \sigma$ and $\tilde\tau$ respectively. Our proposal suggests we need a non-BPS $\tilde{\mathrm{D}}2$-brane and D6-brane inducing changes in the $F_7$-flux and $F_3$-flux respectively. We call their worldvolume tachyons $X_1$ and $X_2$ respectively. The relevant field space metric is
\begin{equation}
    \begin{split}
        \dd s_{\text{field}}^2 = 2 \, \e^{\frac{\tdelta_3}{2} -\frac{\sqrt{3}}{2} \tsigma - \ttau} \dd X_1^2 + 2 \, \e^{\frac{\tdelta_3}{2} - \frac{\sqrt{3}}{2} \tsigma + \ttau} \dd X_2^2 \\
        + \dd \tdelta_3^2 + \dd \tsigma_2 + \dd \ttau^2 & .
    \end{split}
\end{equation}
After the field  rotation
\begin{subequations}
    \begin{align}
        P_{i} & \equiv - \dfrac{1}{2} \biggl( \dfrac{\tdelta_3}{2} - \dfrac{\sqrt{3}}{2} \tsigma + (-1)^i \, \ttau \biggr), \\
        Q & \equiv - \dfrac{1}{2} \biggl( \tdelta_3 + \dfrac{\tsigma}{\sqrt{3}} \biggr),
    \end{align}
\end{subequations}
and defining $P_i = \log(Z_i)$, we find
\begin{equation}
    \dd s_{\text{field}}^2 = 2 \sum_i \frac{\dd X_i^2 + \dd Z_i^2}{Z_i^2} + \dd Q^2,
\end{equation}
i.e. a field space of the form $(\mathbb{H}^2)^2 \times \mathbb{R}$. Then again, one should find how the solutions scale. One can infer that if the $F_7$-flux is given by $N^a$ and the $F_3$-flux by $N^b$, then $\tilde{\Phi}_3 \sim - [3(a+b)/4] \, \log N$, $\tilde{\sigma} \sim [\sqrt{3}(a+b)/4] \, \log N$ and $\tilde{\tau} \sim [(a-b)/2] \, \log N$. With that, the new fields scale as $Z_1 \sim N^{(5a+b)/8}$, $Z_2 \sim N^{(a+5b)/8}$ and $Q \sim [(a+b)/4] \, \log N$. In order to have large volume for the four-torus, we need $a>b$. This guarantees that $X_1 \sim N^a$ scales higher than $Z_1$ in the large $N$-limit. The same happens for $Z_2$ and $X_2 \sim N^b$ when $3b > a > b$, otherwise $X_2$ scales lower than $Z_2$. When $3b>a$, the distance is 
\begin{equation}
    \Delta_1 = \sqrt{\frac{112 a^2 - 13 ab + 32 b^2}{8}} \log N\,.
\end{equation}
In the other case, when $a > 3b$, we have
\begin{equation}
    \Delta_1 = \sqrt{\frac{112 a^2 - 5 ab + 8 b^2}{8}} \log N\,.
\end{equation}
Comparing the tachyon energy densities with the flux potential, we notice that
\begin{equation}
    \frac{V_{X_1}}{V_{\mathrm{AdS}_ \times \mathrm{S}^3\times \mathbb{T}^4}} \sim N^{(b-3a)/4}, \quad \frac{V_{X_2}}{V_{\mathrm{AdS}_ \times \mathrm{S}^3\times \mathbb{T}^4}} \sim N^{(a-3b)/4}.
\end{equation}
Here the potential energy of the D2-tachyon $X_1$ is parametrically smaller than the flux provided by the fluxes (as $a>b$ for the large volume limit). Whether the energy of the D6-tachyon $X_2$ is lower than the flux potential, depends again on whether $a<3b$ or not, similar to how whether $X_2$ scales higher than $Z_2$ and how it influences the distance computation.

\bibliographystyle{apsrev4-1}
\bibliography{references.bib}

\end{document}